\documentclass[12pt,a4paper,fleqn]{article}

\usepackage[tbtags]{amsmath}
\usepackage{amssymb}
\usepackage{mathrsfs}

\newcommand{\ui}{\,\mathrm{i}\;\!}

\newcommand{\uO}{\mathrm{O}}
\newcommand{\xnc}[1]{\hat{#1}}
\newcommand{\ncx}{\hat{x}}
\newcommand{\ncp}{\hat{p}}
\newcommand{\ncpartial}{\hat{\partial}}
\newcommand{\nceth}{\hat{\eth}}
\newcommand{\ncnabla}{\hat{\nabla}}
\newcommand{\opid}{\mathbf{1}}
\newcommand{\opP}{\mathcal{P}}
\newcommand{\opH}{\mathcal{H}}
\newcommand{\opJ}{\mathcal{J}}
\newcommand{\opG}{\mathcal{G}}
\newcommand{\opD}{\mathcal{D}}
\newcommand{\opK}{\mathcal{K}}
\newcommand{\opF}{\mathcal{F}}

\newcommand{\ncopP}{\xnc{\mathcal{P}}}
\newcommand{\ncopH}{\xnc{\mathcal{H}}}
\newcommand{\ncopJ}{\xnc{\mathcal{J}}}
\newcommand{\ncopG}{\xnc{\mathcal{G}}}
\newcommand{\ncopD}{\xnc{\mathcal{D}}}
\newcommand{\ncopK}{\xnc{\mathcal{K}}}

\newcommand{\scopP}{\bar{\mathcal{P}}}
\newcommand{\scopH}{\bar{\mathcal{H}}}
\newcommand{\scopJ}{\bar{\mathcal{J}}}
\newcommand{\scopG}{\bar{\mathcal{G}}}
\newcommand{\scopD}{\bar{\mathcal{D}}}
\newcommand{\scopK}{\bar{\mathcal{K}}}
\renewcommand{\vec}[1]{\mathbf{#1}}
\newcommand{\refcite}[1]{\cite{#1}}
\newcommand{\refscite}[1]{\cite{#1}}

\title{Deformed relativistic and nonrelativistic symmetries on canonical noncommutative spaces}

\author{
Rabin Banerjee\\
  \textit{\small S.N. Bose National Centre for Basic Sciences,}\\[-0.1cm]
  \textit{\small JD Block, Sector 3, Salt Lake, Kolkata 700098, India}\\[-0.1cm]
  {\small E-mail: \texttt{rabin@bose.res.in}}
\and
Kuldeep Kumar\\
  \textit{\small Department of Physics, Panjab University,
  Chandigarh 160014, India}\\[-0.1cm]
  {\small E-mail: \texttt{kuldeepk@pu.ac.in}}
}

\date{}

\begin{document}

\maketitle

\begin{abstract}
We study the general deformed conformal-Poincar\'e (Galilean) symmetries consistent with relativistic (nonrelativistic) canonical noncommutative spaces. In either case we obtain deformed generators, containing arbitrary free parameters, which close to yield new algebraic structures. We show that a particular choice of these parameters reproduces the undeformed algebra. The modified coproduct rules and the associated Hopf algebra are also obtained. Finally, we show that for the choice of parameters leading to the undeformed algebra, the deformations are represented by twist functions. {\footnotesize [Journal reference: \textit{Phys.\ Rev.}\ D 75 (2007) 045008]}
\end{abstract}

\bigskip


\section{\label{sec:deform-intro}Introduction}

In a series of papers Wess \cite{Wess:2003da} and collaborators \cite{Dimitrijevic:2004rf, Aschieri:2005yw, Koch:2004ud} have discussed the deformation of various symmetries on noncommutative spaces. The modified coproduct rule obtained for the Poincar\'e generators is found to agree with an alternative quantum-group-theoretic derivation \cite{Chaichian:2004za, Chaichian:2004yh, Matlock:2005zn} based on the application of twist functions \cite{Oeckl:2000eg}. The extension of these ideas to field theory and possible implications for Noether symmetry are discussed in \refscite{Banerjee:2004ev, Gonera:2005hg, Calmet:2004ii}. An attempt to extend such notions to supersymmetry has been done in \refscite{Kobayashi:2004ep, Zupnik:2005ut, Ihl:2005zd, Banerjee:2005ig}. Recently, the deformed Poincar\'e generators for  Lie-algebraic $\theta$ (rather than a constant $\theta$) \cite{Lukierski:2005fc} and Snyder \cite{Snyder:1946qz} noncommutativity \cite{Banerjee:2006wf} have also been analysed.

In this paper we develop an algebraic method for analysing the deformed relativistic and nonrelativistic symmetries in noncommutative spaces with a constant noncommutativity parameter. By requiring the twin conditions of consistency with the noncommutative space and  closure of the Lie algebra, we obtain deformed generators with arbitrary free parameters. For conformal-Poincar\'e symmetries we show that a specific choice of these parameters yields the undeformed algebra, although the generators are still deformed. For the nonrelativistic (Schr\"odinger \cite{Niederer:1972, Hagen:1972pd, Burdet:1972xd}) case two possibilities are discussed for introducing the free parameters. In one of these there is no choice of the parameters that yields the undeformed algebra.

A differential-operator realisation of the deformed generators is given in the coordinate and momentum representations. The various expressions naturally contain the free parameters. For the particular choice of these parameters that yields the undeformed algebra, the deformations in the generators drop out completely in the momentum representation.

The modified comultiplication rules (in the coordinate representation) and the associated Hopf algebra are calculated. For the choice of parameters that leads to the undeformed algebra these rules agree with those obtained by an application of the abelian twist function on the primitive comultiplication rule. (For the conformal-Poincar\'e case this computation of modified coproduct rules using the twist function already exists in the literature \cite{Chaichian:2004za, Chaichian:2004yh, Matlock:2005zn, Oeckl:2000eg}, but a similar analysis for the nonrelativistic symmetries is new and presented here.) For other choices of the free parameters the deformations cannot be represented by twist functions. The possibility that there can be such deformations  also arises in the context of $\kappa$-deformed symmetries \cite{Lukierski:2006fv}.
  

\section{\label{sec:deform-conf-rel}Deformed conformal-Poincar\'e algebra}

Here we analyse the deformations in the full conformal-Poincar\'e generators compatible with a canonical (constant) noncommutative spacetime. We find that there exists a one-parameter class of deformed special conformal generators that yields a closed algebra whose structure is completely new. A particular value of the parameter leads to the undeformed algebra.

We begin by presenting an algebraic approach whereby compatibility is achieved with noncommutative spacetime by the various Poincar\'e generators. This spacetime is characterised by the algebra
\begin{equation}\label{brac:nc101}
\big[ \ncx^{\mu}, \ncx^{\nu} \big] = \ui \theta^{\mu \nu}, \qquad
\big[ \ncp_{\mu}, \ncp_{\nu} \big] = 0, \qquad
\big[ \ncx^{\mu}, \ncp_{\nu} \big] = \ui {\delta^{\mu}}_{\nu}.
\end{equation}
It follows that, for any spacetime transformation,
\begin{equation}\label{nc102}
\big[ \delta \ncx^{\mu}, \ncx^{\nu} \big] + \big[ \ncx^{\mu}, \delta \ncx^{\nu} \big] = 0
\end{equation}
for constant $\theta$. Translations, $\delta \ncx^{\mu} = a^{\mu}$, with constant $a^{\mu}$, are obviously compatible with \eqref{nc102}. The generator of the transformation consistent with $\delta \ncx^{\mu} = \ui a^{\sigma} [ \ncopP_{\sigma}, \ncx^{\mu} ]$ is $\ncopP_{\mu} = \ncp_{\mu}$. For a Lorentz transformation, $\delta \ncx^{\mu} = \omega^{\mu \nu} \ncx_{\nu}$, $\omega^{\mu \nu} = - \omega^{\nu \mu}$, the requirement \eqref{nc102} implies ${\omega^\mu}_\lambda \theta^{\lambda \nu} - {\omega^\nu}_\lambda \theta^{\lambda \mu} = 0$, which is not satisfied except for two dimensions. Therefore, in general, the usual Lorentz transformation is not consistent with \eqref{nc102}. A deformation of the Lorentz transformation is therefore mandatory. We consider the minimal $\uO(\theta)$ deformation:
\begin{equation*}\label{nc104}
\delta \ncx^{\mu} = \omega^{\mu \nu} \ncx_{\nu} + n_1 {\omega^{\mu}}_{\nu} \theta^{\nu \sigma} \ncp_{\sigma} + n_2 {\omega_{\nu}}^{\sigma} \theta^{\mu \nu} \ncp_{\sigma} + n_3 \omega_{\nu \sigma} \theta^{\nu \sigma} \ncp^{\mu},
\end{equation*}
where $n_1$, $n_2$ and $n_3$ are coefficients to be determined by consistency arguments. Now the  generator,
\begin{equation*}\label{nc105}
\ncopJ^{\mu \nu}
= \ncx^{\mu} \ncp^{\nu} + \lambda_{1} \theta^{\mu \sigma} \ncp_{\sigma} \ncp^{\nu} + \tfrac{1}{2} \lambda_{2} \theta^{\mu \nu} \ncp^{2} - \langle \mu \nu \rangle,
\end{equation*}
where $\langle \mu \nu \rangle$ denotes the preceding terms with $\mu$ and $\nu$ interchanged, is consistent with $\delta \ncx^{\mu} = - (\ui/2) \omega_{\rho \sigma} [ \ncopJ^{\rho \sigma}, \ncx^{\mu} ]$ for $n_1 = n_2 + 1 = \lambda_1$, $n_3 = -\lambda_2$, a result which follows on using \eqref{brac:nc101}. It is therefore clear that $n_1 = n_2 = 0$ is not possible, which necessitates the modification of the transformation as well as the generator. It turns out that
\begin{equation*}\label{nc106}
\big[ \ncopJ^{\mu \nu}, \ncopJ^{\rho \sigma} \big]
= \ui \big[ \eta^{\mu \rho} \ncopJ^{\nu \sigma} - \theta^{\mu \rho} \big\{ (2\lambda_1 - 1) \ncp^{\nu} \ncp^{\sigma} + \lambda_2 \ncp^2 \eta^{\nu \sigma} \big\} \big] - \langle \mu \nu \rho \sigma \rangle,
\end{equation*}
with the notation $Z^{\cdots} - \langle \mu \nu \rho \sigma \rangle \equiv \left( Z^{\cdots} - \langle \mu \nu \rangle \right) - \langle  \rho \sigma \rangle$. The closure of the normal Lorentz algebra is obtained only for $\lambda_1 = \frac{1}{2}$ and $\lambda_2 = 0$ \cite{Koch:2004ud}.

Similarly, the usual scale transformation, $\delta \ncx^{\mu} =  \alpha \ncx^{\mu}$, is not consistent with \eqref{nc102}. A minimally deformed form of the transformation is $\delta \ncx^{\mu} = \alpha \ncx^{\mu} + \alpha n \theta^{\mu \nu} \ncp_{\nu}$. The consistency, $\delta \ncx^{\mu} = \ui \alpha [ \ncopD, \ncx^{\mu} ]$, is achieved only for $n = 1$ by $\ncopD = \ncx^\mu \ncp_\mu$. Likewise, starting with the minimally deformed form of the special conformal transformation we find that the  generator,
\begin{equation}\label{nc110}
\ncopK^{\rho} = 2 \ncx^{\rho} \ncx^{\sigma} \ncp_{\sigma} - \ncx^2 \ncp^\rho + \eta_1 \theta^{\rho \sigma} \ncp_{\sigma} + \eta_2 \theta^{\rho \sigma} \ncx^{\beta} \ncp_{\beta} \ncp_{\sigma} + \eta_3 \theta^{\sigma \beta} \ncx^{\sigma} \ncp_{\beta} \ncp^{\rho}
\end{equation}
is consistent with $\delta \ncx^{\mu} = \ui \omega_{\rho} [ \ncopK^{\rho}, \ncx^{\mu} ]$ for $\eta_2 = \eta_3 = 0$.

This completes our demonstration of the compatibility of the various transformation laws with the basic noncommutative algebra. However, achieving consistency with the transformation and closure of the algebra are two different things. It turns out that the minimal $\uO(\theta)$ deformation, while preserving consistency, does not yield a closed algebra. Indeed we find that $[ \ncopK^\rho, \ncopD ]$ algebra does not close, necessitating the inclusion of $\uO(\theta^2)$ terms in the deformed transformation and the deformed generator. An appropriately deformed form,
\begin{equation*}\label{nc113}
\begin{split}
\ncopK^{\rho}
&= [\text{right-hand side of Eq.~\eqref{nc110}}] \\
&\quad\, {} + \eta_4 \theta^{\alpha \beta} {\theta_{\alpha}}^{\sigma} \ncp_{\sigma} \ncp_{\beta} \ncp^{\rho} + \eta_5 \theta^{\alpha \beta} \theta_{\alpha \beta} \ncp^2 \ncp^{\rho} + \eta_6 \theta^{\rho \alpha} {\theta_{\alpha}}^{\sigma} \ncp_{\sigma} \ncp^2,
\end{split}
\end{equation*}
involves 6 free parameters. However, the closure of the $[ \ncopK^\rho, \ncopD ]$ algebra fixes 5 parameters, $\eta_2 = -\eta_3 = -4 \eta_4 = 1$, $\eta_5 = \eta_6 = 0$, leaving only one, $\eta_1$, as free. The final form of the deformed generators,
\begin{equation}\label{nc115}
\begin{aligned}
&\ncopP_{\mu} = \ncp_{\mu}, \qquad
\ncopJ^{\mu \nu} = \ncx^{\mu} \ncp^{\nu} + \tfrac{1}{2} \theta^{\mu \sigma} \ncp_{\sigma} \ncp^{\nu} - \langle \mu \nu \rangle, \qquad
\ncopD = \ncx^\mu \ncp_\mu, \\
&\ncopK^{\rho} = 2 \ncx^{\rho} \ncx^{\sigma} \ncp_{\sigma} - \ncx^2 \ncp^\rho + \eta_1 \theta^{\rho \sigma} \ncp_{\sigma} + \theta^{\rho \sigma} \ncx^{\beta} \ncp_{\beta} \ncp_{\sigma} - \theta^{\sigma \beta} \ncx_{\sigma} \ncp_{\beta} \ncp^{\rho} - \tfrac{1}{4} \theta^{\alpha \beta} {\theta_{\alpha}}^{\sigma} \ncp_{\sigma} \ncp_{\beta} \ncp^{\rho},
\end{aligned}
\end{equation}
involves one free parameter. The algebra satisfied by the generators is such that the Poincar\'e sector remains unaffected changing only the conformal sector:
\begin{equation}\label{nc116}
\begin{aligned}
&\big[ \ncopD, \ncopP^{\mu} \big] = \ui \ncopP^{\mu}, \qquad
\big[ \ncopD, \ncopJ^{\mu \nu} \big]  = 0, \qquad
\big[ \ncopK^{\rho}, \ncopP^{\mu} \big] = 2 \ui \big( \eta^{\rho \mu} \ncopD + \ncopJ^{\rho \mu} \big), \\
&\big[ \ncopK^{\rho}, \ncopJ^{\mu \nu} \big] = - \ui \big[ \eta^{\rho \mu} \ncopK^{\nu} + (\ui + \eta_1) \big( \theta^{\rho \mu} \ncopP^{\nu} - \eta^{\rho \mu} \theta^{\nu \sigma} \ncopP_{\sigma} \big) \big] - \langle \mu\nu \rangle, \\
&\big[ \ncopK^{\rho}, \ncopD \big] = \ui \big[ \ncopK^{\rho} - 2 (\ui + \eta_1) \theta^{\rho \mu} \ncopP_{\mu} \big], \\
&\big[ \ncopK^{\rho}, \ncopK^{\mu} \big] = -2 \ui (\ui + \eta_1) \big( \theta^{\rho \mu} \ncopD - \theta^{\mu \sigma} {\ncopJ^{\rho}{}}_{\sigma} \big) - \langle \rho\mu \rangle.
\end{aligned}
\end{equation}
A one-parameter class of closed algebras is found. Fixing $\eta_1 = - \ui$ yields the usual (undeformed) Lie algebra. In that case the deformed special conformal generator also agrees with the result given in \refcite{Banerjee:2005ig}.

The relations in \eqref{brac:nc101} are easily reproduced by representing
\begin{equation*}\label{repr-x}
\ncx^\mu = \ncx^\mu, \qquad
\ncp_\mu = - \ui \ncpartial_\mu \equiv -\ui \frac{\partial}{\partial \ncx^\mu}.
\end{equation*}
In this coordinate representation, the generators read
\begin{equation}\label{nc-cor115}
\begin{aligned}
&\ncopP_{\mu} = -\ui \ncpartial_{\mu}, \qquad
\ncopJ^{\mu \nu} = -\ui \ncx^{\mu} \ncpartial^{\nu} - \tfrac{1}{2} \theta^{\mu \sigma} \ncpartial_{\sigma} \ncpartial^{\nu} - \langle \mu \nu \rangle, \qquad
\ncopD = -\ui \ncx^\mu \ncpartial_\mu, \\
&\begin{aligned}
\ncopK^{\rho} &= -2 \ui \ncx^{\rho} \ncx^{\sigma} \ncpartial_{\sigma} + \ui \ncx^2 \ncpartial^\rho - \ui  \eta_1 \theta^{\rho \sigma} \ncpartial_{\sigma} - \theta^{\rho \sigma} \ncx^{\beta} \ncpartial_{\beta} \ncpartial_{\sigma} + \theta^{\sigma \beta} \ncx_{\sigma} \ncpartial_{\beta} \ncpartial^{\rho} \\
&\quad\, {} - \frac{\ui}{4} \theta^{\alpha \beta} {\theta_{\alpha}}^{\sigma} \ncpartial_{\sigma} \ncpartial_{\beta} \ncpartial^{\rho}.
\end{aligned}
\end{aligned}
\end{equation}
One may also choose the momentum representation,
\begin{equation*}
\ncp_{\mu} = \ncp_{\mu}, \qquad
\ncx^{\mu} = \ui \nceth^{\mu} - \frac{1}{2} \theta^{\mu \nu} \ncp_{\nu} \equiv \ui \frac{\partial}{\partial \ncp_{\mu}} - \frac{1}{2} \theta^{\mu \nu} \ncp_{\nu},
\end{equation*}
which gives
\begin{gather*}
\ncopP^{\mu} = \ncp^{\mu}, \qquad
\ncopJ^{\mu \nu} = \ui \big( \ncp^\nu \nceth^\mu - \ncp^\mu \nceth^\nu \big), \qquad
\ncopD = \ui \ncp_{\mu} \nceth^\mu + \ui N, \\
\ncopK^{\rho} = \ncp^\rho \nceth^2 - 2 \ncp_\sigma \nceth^\rho \nceth^\sigma - 2 N \nceth^\rho + (\eta_1 + \ui) \theta^{\rho \sigma} \ncp_\sigma,
\end{gather*}
where $N = {\delta^\mu}_\mu$ is the number of spacetime dimensions. For $\eta_1 = -\ui$, when the generators satisfy the undeformed algebra, the deformation in $\ncopK^\rho$ drops out and all the generators in momentum representation have exactly the same structure as in the commutative description.

The deformed generators lead to new comultiplication rules. The coproduct rules for the Poincar\'e sector were earlier derived in \refscite{Wess:2003da, Koch:2004ud, Chaichian:2004za, Chaichian:2004yh} and for the conformal sector in \refscite{Matlock:2005zn, Banerjee:2005ig}. The free parameter appearing in $\ncopK^{\rho}$ does not appear explicitly in the coproduct $\Delta (\ncopK^{\rho})$. Computing the basic Hopf algebra, it turns out that the Hopf algebra can be read off from \eqref{nc116} by just replacing the generators by the coproducts. For example,
\begin{equation*}
\!\big[ \Delta ( \ncopK^{\rho} ), \Delta ( \ncopD ) \big] = \ui \big[ \Delta ( \ncopK^{\rho} ) - 2 (\ui + \eta_1) \theta^{\rho \mu} \Delta ( \ncopP_{\mu} ) \big].
\end{equation*}


\section{\label{sec:deform-conf-nonrel}Deformed Schr\"odinger and conformal-Galilean algebras}

Now we consider separately the Schr\"odinger symmetry and the conformal-Galilean symmetry, both of which are extensions of the Galilean symmetry.

The standard Schr\"odinger algebra is given by extending the Galilean algebra, which involves Hamiltonian ($\opH$), translations ($\opP^{i}$), rotations ($\opJ^{ij}$) and boosts ($\opG^{i}$), with the algebra of dilatation ($\opD$) and expansion or special conformal transformation ($\opK$). The standard free-particle representation of this algebra is given by
\begin{gather*}
\opH = \frac{1}{2m} \vec{p}^{2}, \qquad
\opP^{i} = p^{i}, \qquad
\opJ^{ij} = x^{i} p^{j} - x^{j} p^{i}, \qquad
\opG^{i} = m x^{i} - t p^{i}, \\
\opD = p^i x^i - \frac{t}{m} \vec{p}^2, \qquad \opK = \frac{m}{2} \left( \vec{x} - \frac{t}{m} \vec{p} \right)^2.
\end{gather*}

Now we introduce noncommutativity in space:
\[
[ \ncx^{i}, \ncx^{j} ] = \ui \theta^{ij}, \qquad
[ \ncp^{i}, \ncp^{j} ] = 0, \qquad
[ \ncx^{i}, \ncp^{j} ] = \ui \delta^{ij}.
\]
Like the deformed conformal-Poincar\'e case, we follow a two-step algebraic process. First, by requiring the compatibility of transformations with the noncommutative space, a general deformation of the generators is obtained. A definite structure emerges after demanding the closure of the algebra. The linear momentum and the Hamiltonian retain their original forms because the algebra of $\ncp^{i}$ is identical to $p^{i}$. For other generators we consider the minimal deformation. The final form of the generators,
\begin{equation}\label{gen-nc2:gal}
\begin{aligned}
&\ncopH = \frac{1}{2m} \vec{\ncp}^{2}, \qquad
\ncopP^{i} = \ncp^{i}, \qquad
\ncopJ^{ij} = \ncx^{i} \ncp^{j} + \tfrac{1}{2} \theta^{ik} \ncp^{k} \ncp^{j} + \tfrac{1}{2} \lambda_{2} \theta^{ij} \vec{\ncp}^{2} - \langle ij \rangle, \\
&\ncopG^{i} = m \ncx^{i} - t \ncp^{i} + \lambda_{3} m \theta^{ij} \ncp^{j},\qquad
\ncopD = \ncp^i \ncx^i - \frac{t}{m} \vec{\ncp}^2, \\
&\ncopK = \frac{m}{2} \left( \vec{\ncx} - \frac{t}{m} \vec{\ncp} \right)^2 + \frac{m}{4} \theta^{ij} \ncx^i \ncp^j,
\end{aligned}
\end{equation}
leads to a nonstandard closure of the algebra:
\begin{equation}\label{alg-nc:gal-11}
\begin{aligned}
&\big[ \ncopJ^{ij}, \ncopJ^{k\ell} \big] = \ui \big( \delta^{ik} \ncopJ^{j\ell} - 2 \theta^{ik} \lambda_2 m \ncopH \delta^{j\ell} \big) - \langle i j k \ell \rangle, \\
&\big[ \ncopG^i, \ncopG^j \big] = \ui m^2 (1- 2\lambda_3) \theta^{ij}, \\
&\big[ \ncopG^i, \ncopJ^{jk} \big] = \ui \Big[ \delta^{ik} \ncopG^j + m \left( \tfrac{1}{2} - \lambda_3 \right) \big( \theta^{ij} \ncopP^k + \delta^{ik} \theta^{jm} \ncopP^m \big) + m \lambda_2 \theta^{jk} \ncopP^i \Big] - \langle j k \rangle, \\
&\big[ \ncopJ^{ij}, \ncopD \big] = - 4 \ui m \lambda_2 \theta^{ij} \ncopH, \qquad
\big[ \ncopD, \ncopG^i \big] = -\ui \left[ \ncopG^i + m (1 - 2 \lambda_3) \theta^{ij} \ncopP^j \right], \\
&\big[ \ncopK, \ncopP^i \big] = \ui \left[ \opG^i + \left( \tfrac{1}{4} - \lambda_3 \right) m \theta^{ij} \ncopP^j \right], \\
&\big[ \ncopJ^{ij}, \ncopK \big] = \ui \left[ \frac{m}{4} \big( \theta^{ik} \ncopJ^{kj} - \theta^{jk} \ncopJ^{ki} \big) - 2 \lambda_2 m \theta^{ij} \ncopD \right], \\
&\big[ \ncopK, \ncopG^i \big] = -\ui m \theta^{ij} \left[ \left( \tfrac{3}{4} - \lambda_3 \right) \ncopG^j + \left( \lambda_3^2 - \lambda_3 + \tfrac{1}{4} \right) m \theta^{jk} \ncopP^k \right], \\
&\big[ \ncopD, \ncopK \big] = -\ui \left( 2 \ncopK + \frac{m}{4} \theta^{ij} \ncopJ^{ij} - \tfrac{1}{2} \lambda_2 m^2 \theta^{ij} \theta^{ij} \ncopH \right),
\end{aligned}
\end{equation}
the other brackets remaining unaltered. 

Some comments are in order. We have obtained the deformed Schr\"odinger algebra involving two parameters, $\lambda_2$ and $\lambda_3$. For $\theta\rightarrow 0$, the deformed algebra reduces to the undeformed one. A distinctive feature is that there is no choice of the free parameters for which the standard (undeformed) algebra can be reproduced. This is an obvious and important difference from the Poincar\'e treatment.

It is however possible to obtain an alternative deformation which, for a particular choice of parameters, yields the undeformed algebra. We notice that the brackets involving all generators other than $\ncopK$ reduce to the standard ones by fixing $\lambda_2 = 0$ and $\lambda_3 = \frac{1}{2}$, although the generators are deformed. So we allow $\uO(\theta^2)$ terms in $\ncopK$. Demanding the closure of $[ \ncopH, \ncopK ]$ and $[ \ncopD, \ncopK ]$ brackets yields $\uO(\theta^2)$-deformed $\ncopK$:
\begin{equation}\label{gen-nc2:K-2}
\ncopK = \frac{m}{2} \left( \vec{\ncx} - \frac{t}{m} \vec{\ncp} \right)^2 + \lambda_6 m \theta^{ij} \ncx^i \ncp^j + m \left( \frac{1}{8} - \frac{\lambda_6}{2} \right) \theta^{ij} \theta^{jk} \ncp^i \ncp^k.
\end{equation}
The brackets involving this $\ncopK$ turn out to be
\begin{equation}\label{alg-nc:K-2}
\begin{aligned}
&\big[ \ncopH, \ncopK \big] = -\ui \ncopD, \qquad
\big[ \ncopK, \ncopP^i \big] = \ui \big[ \opG^i + \big( \lambda_6 - \lambda_3 \big) m \theta^{ij} \ncopP^j \big], \\
&\big[ \ncopJ^{ij}, \ncopK \big] = \ui \left[ \left( \tfrac{1}{2} - \lambda_6 \right) m \big( \theta^{ik} \ncopJ^{kj} - \theta^{jk} \ncopJ^{ki} \big) - 2 \lambda_2 m \theta^{ij} \ncopD \right], \\
&\big[ \ncopK, \ncopG^i \big] = -\ui m \theta^{ij} \Big[ \left( 1 - \lambda_3 - \lambda_6 \right) \ncopG^j + \left( \lambda_3^2 - \lambda_3 + \tfrac{1}{4} \right) m \theta^{jk} \ncopP^k \Big], \\
&\big[ \ncopD, \ncopK \big] = \ui \left[ - 2 \ncopK + \left( \tfrac{1}{2} - \lambda_6 \right) m \theta^{ij} \big( 2 \lambda_2 m \theta^{ij} \ncopH - \ncopJ^{ij} \big) \right]
\end{aligned}
\end{equation}
which give us another deformed Schr\"odinger algebra involving three parameters, $\lambda_2$, $\lambda_3$ and $\lambda_6$. It is easily seen from \eqref{alg-nc:K-2} that the particular choice of parameters, $\lambda_2 = 0$ and $\lambda_3 = \lambda_6 = \frac{1}{2}$, reproduces the standard algebra. This agrees with \refcite{Banerjee:2005zt}. Now onwards we shall restrict to $\ncopK$ given by \eqref{gen-nc2:K-2} whenever expansions are considered.

In the coordinate representation,
\begin{gather*}
\ncopD = - \ui \ncx^i \ncpartial^i + \frac{t}{m} \vec{\ncnabla}^2 - \ui N', \\
\ncopK = \frac{m}{2} \vec{\ncx}^2 - \frac{t^2}{2m} \vec{\ncnabla}^2 + \ui t \ncx^i \ncpartial^i + \ui \frac{tN'}{2} - \ui \lambda_6 m \theta^{ij} \ncx^i \ncpartial^j - m \left( \frac{1}{8} - \frac{\lambda_6}{2} \right) \theta^{ij} \theta^{jk} \ncpartial^i \ncpartial^k,
\end{gather*}
where $N' = \delta^{ii}$ is the number of space dimensions. The momentum representation of $\ncopD$ is $\ncopD = \ui \ncp^i \nceth^i - t\vec{\ncp}^2/m$. The representation for $\ncopK$ involves a deformation which, expectedly, drops out for $\lambda_6 = \frac{1}{2}$ that corresponds to the standard algebra.

The comultiplication rules, using the coordinate representation, turn out to be
\begin{gather}
\label{co-ncH}
\Delta ( \ncopH ) = \ncopH \otimes \opid + \opid \otimes \ncopH + \frac{1}{m} \ncopP^i \otimes \ncopP^i,\\
\label{co-ncP}
\Delta ( \ncopP^i ) = \ncopP^i \otimes \opid + \opid \otimes \ncopP^i, \\
\label{co-ncJ}
\begin{split}
\Delta ( \ncopJ^{ij} )
&= \tfrac{1}{2} \big[ \ncopJ^{ij} \otimes \opid + \opid \otimes \ncopJ^{ij} + \theta^{im} \big( \ncopP^j \otimes \ncopP^m - \ncopP^m \otimes \ncopP^j \big) \big] \\
&\quad\, {} + \lambda_2 \theta^{ij} \ncopP^m \otimes \ncopP^m - \langle i j \rangle,
\end{split} \displaybreak[0]\\
\label{co-ncG}
\begin{split}
\Delta ( \ncopG^i ) &= \tfrac{1}{2} \big[ \ncopG^i \otimes \opid + \opid \otimes \ncopG^i - t \big( \ncopP^i \otimes \opid + \opid \otimes \ncopP^i \big) \\
&\qquad\, {} + m \theta^{ij} \big\{ (\lambda_3 - 1) \ncopP^j \otimes \opid + \lambda_3 \opid \otimes \ncopP^j \big\} \big],
\end{split} \displaybreak[0]\\
\label{co-ncD}
\Delta ( \ncopD ) =  \ncopD \otimes \opid + \opid \otimes \ncopD - \frac{2t}{m} \ncopP^i \otimes \ncopP^i + \ui \frac{N'}{2} \opid \otimes \opid + \theta^{ij} \ncopP^i \otimes \ncopP^j, \displaybreak[0]\\
\label{co-ncK}
\begin{split}
\Delta ( \ncopK )
&=  \ncopK \otimes \opid + \opid \otimes \ncopK - \ui \frac{tN'}{2} \opid \otimes \opid \\
&\quad\, {} + \frac{1}{2m} \big[ t^2 \ncopP^i \otimes \ncopP^i - \ncopG^i \otimes \ncopG^i - t \big( \ncopP^i \otimes \ncopG^i + \ncopG^i \otimes \ncopP^i \big) \big] \\
&\quad\, {} - \tfrac{1}{2} \theta^{ij} \big[ t \ncopP^i \otimes \ncopP^j + (\lambda_3 - 1) \ncopP^i \otimes \ncopG^j + \lambda_3 \ncopG^j \otimes \ncopP^i \big] \\
&\quad\, {} - \frac{m}{2} \theta^{ij} \theta^{ik} \left( \lambda_3^2 - \lambda_3 + \tfrac{1}{2} \right) \ncopP^j \otimes \ncopP^k.
\end{split}
\end{gather}
Among the free parameters $\lambda_2$, $\lambda_3$ and $\lambda_6$ appearing in the definition of the deformed generators, only the first two occur in the expressions for the deformed coproducts. The parameter $\lambda_6$, which is present in $\ncopK$, however, does not occur in $\Delta(\ncopK)$. Expectedly, it turns out that the Hopf algebra can be directly read off from the algebra (Eqs. \eqref{alg-nc:gal-11}, etc.) by just replacing the generators by the coproducts.

There is an alternative method, based on quantum-group-theoretic arguments, of computing the coproducts \cite{Chaichian:2004za, Chaichian:2004yh, Matlock:2005zn}. This is obtained for the particular case when the deformed generators satisfy the undeformed algebra. In our analysis it corresponds to the choice $\lambda_2 = 0$, $\lambda_3 = \lambda_6 = \frac{1}{2}$. The essential ingredient is the application of the abelian twist function,
\[
\opF = \exp \left( \frac{\ui}{2} \theta^{ij} \opP^i \otimes \opP^j \right),
\]
as a similarity transformation on the primitive coproduct rule to abstract the deformed rule. After some calculations it can be shown that the deformed coproduct rule \eqref{co-ncG}, for example, for the specific values of the free parameters already stated, is obtained by identifying
\begin{equation*}\label{co-nctG}
\Delta ( \ncopG^i ) = \big[ \opF {\,} \Delta ( \opG^i ) {\,} \opF^{-1} \big]_{\opG^i \rightarrow \ncopG^i, \opP^j \rightarrow \ncopP^j}.
\end{equation*}
The coproducts for other generators can similarly be obtained from the same twist element.

Strictly speaking, the algebra obtained by enlarging the Galilean algebra by including dilatations and expansions is not a conformal algebra since it does not inherit some basic characteristics like vanishing of the mass, equality of the number of translations and the special conformal transformations, etc. However, since it is a symmetry of the Schr\"odinger equation, this enlargement of the Galilean algebra is appropriately referred to as the Schr\"odinger algebra. It is possible to discuss the conformal extension of the Galilean algebra by means of a nonrelativistic contraction of the relativistic conformal-Poincar\'e algebra. Recently this was discussed for the particular case of three dimensions \cite{Lukierski:2005xy}. This algebra is different from the Schr\"odinger algebra discussed earlier. We scale the generators and the noncommutativity parameter as
\begin{equation}
\begin{aligned}
&\ncopD = \scopD, \qquad
\ncopK^{\rho} = \big( \ncopK^0, \ncopK^i \big) = \Big( c \scopK, c^2 \scopK^i \Big), \qquad
\ncopP^{\mu} = \big( \ncopP^0, \ncopP^i \big) = \Big( \scopH /c, \scopP^i \Big), \\
&\ncopJ^{\mu \nu} = \big( \ncopJ^{0i}, \ncopJ^{ij} \big) = \Big( c \scopG^i, \scopJ^{ij} \Big), \qquad
\theta^{\mu \nu} = \big( \theta^{0i}, \theta^{ij} \big) = \Big( 0, c^2 \bar{\theta}^{ij} \Big),
\end{aligned}
\end{equation}
where $c$ is the velocity of light. We use this scaling in \eqref{nc116} and take the limit $c \rightarrow \infty$. Finally we redefine to choose the same symbols for the nonrelativistic case ($\scopD \rightarrow \ncopD$, etc.). Then we get the deformed algebra
\begin{equation}
\begin{aligned}
&\big[ \ncopD, \ncopH \big] = \ui \ncopH, \qquad
\big[ \ncopD, \ncopP^i \big] = \ui \ncopP^i, \qquad
\big[ \ncopD, \ncopJ^{ij} \big] = 0, \qquad
\big[ \ncopD, \ncopG^i \big] = 0, \\
&\big[ \ncopK, \ncopH \big] = 2 \ui \eta^{00} \ncopD, \qquad
\big[ \ncopK, \ncopP^i \big] = 2 \ui \ncopG^i, \qquad
\big[ \ncopK, \ncopD \big] = \ui \ncopK, \qquad
\big[ \ncopK, \ncopK \big] = 0, \\
&\big[ \ncopK, \ncopG_i \big] = - \ui \eta^{00} \big[ \ncopK^i - (\ui + \eta_1) \theta^{ij} \ncopP_j \big], \qquad
\big[ \ncopK, \ncopK^i \big] = 2 \ui (\ui + \eta_1) \theta^{ij} \ncopG_j, \\
&\big[ \ncopK, \ncopJ^{ij} \big] = 0, \qquad
\big[ \ncopK^i, \ncopH \big] = - 2 \ui \ncopG^i, \qquad
\big[ \ncopK^i, \ncopP^j \big] = 0, \qquad
\big[ \ncopK^i, \ncopG^j \big] = 0, \\
&\big[ \ncopK^i, \ncopJ^{jk} \big] = - \ui \big[ \eta^{ij} \ncopK^k + (\ui + \eta_1) \big( \theta^{ij} \ncopP^k - \eta^{ij} \theta^{k\ell} \ncopP_\ell \big) \big] - \langle j k \rangle, \\
&\big[ \ncopK^i, \ncopD \big] = \ui \big[ \ncopK^i - 2 (\ui + \eta_1) \theta^{ij} \ncopP_j \big],
\end{aligned}
\end{equation}
which also contains a free parameter. Restricting to three dimensions and the specific choice $\eta_1 = -\ui$ reproduces the results obtained recently in \refcite{Lukierski:2005xy}.


\section{\label{sec:deform-conlu}Conclusions}

We have analysed the deformed conformal-Poincar\'e, Schr\"odinger and conformal-Galilean symmetries compatible with the canonical (constant) noncommutative spacetime and found new algebraic structures.

For the conformal-Poincar\'e case we found a one-parameter class of deformed special conformal generators that yielded a closed algebra whose structure was completely new. Fixing the arbitrary parameter reproduced the usual (undeformed) Lie algebra.

Next we considered the Schr\"odinger symmetry. Here we obtained the deformed Schr\"odinger algebra involving two parameters. The closure of this algebra yielded new structures. The generators involved $\uO(\theta)$ deformations. For $\theta\rightarrow 0$, the deformed algebra reduced to the undeformed one. However, a distinctive feature was that there was no choice of the free parameters for which the standard (undeformed) algebra could be reproduced. Exploring other possibilities, then we obtained an alternative deformation which, for a particular choice of parameters, indeed reproduced the undeformed algebra. In this case the modified special conformal generator involved $\uO(\theta^2)$ terms. The deformed Schr\"odinger algebra now involved three parameters, a particular choice of which reproduced the standard algebra.

Finally we discussed the conformal extension of the Galilean algebra by means of a nonrelativistic contraction of the relativistic conformal-Poincar\'e algebra. This algebra is different from the Schr\"odinger algebra, both in the commutative and noncommutative descriptions. The present analysis can be extended to other (nonconstant) types of noncommutativity. Some results in this direction have already been provided for the Snyder space \cite{Banerjee:2006wf}.


\section*{Acknowledgement}

K.K. thanks the Council of Scientific and Industrial Research, Government of India, for financial support during the period when the major part of this work was completed at S.N. Bose National Centre for Basic Sciences, Kolkata.



\end{document}